\newenvironment{centre}{\begin{center}}{\end{center}}\documentstyle[12pt]{article}
\newlength\Cscr\newlength\Csave\newlength
\Ctenthex\newlength\CFxsize\newlength
\CFxsizeps\newlength\CFsizemakebox\newlength\CFleftcrop\newlength
\CFrightcrop\newlength\CZtbldist\newlength\CZfigdist\newlength\CGDnum
\newlength\CGDtext\newcounter{Cscr}\newcounter
{CBcit}\newcounter{CFlette%
r}\newcounter{Ceqin%
dent}\newcounter{CBtnc}\newcounter{CBtntc}%
\newcounter{CEht}\newcounter{Cbscurr}%
\newcounter{CbsB}\newcounter{CbsC}\newcounter{C%
bsD}
\hoffset\Cscr
\voffset\Cscr
\protect
\begin{document}\renewcommand\theequation{\arabic{equati%
on}}\renewcommand\thetable{\arabic{table}}\renewcommand\thefigure
{\arabic{figure}}\renewcommand\thesection{\arabic{section}}%
\renewcommand\thesubsection{\arabic{section}.\arabic{subsection}%
}\renewcommand\thesubsubsection{\arabic{section}.\arabic{subsect%
ion}.\arabic{subsubsection}}\setcounter{CEht}{10}\setcounter{Cbs%
A}{1}\setcounter{CbsB}{1}\setcounter{CbsC}{1}\setcounter{CbsD}{1%
}\setlength{\CFxsize}{0.998\hsize}\hfill UM--P--95/51\par{%
\centering\protect\mbox{}\\*[\baselineskip]{\large\bf$[p,q]\neq i%
\hbar$}\\*}\addtocounter{CBtntc}{1}{\centering\protect\mbox{}\\J%
ohn P.~Costella$^{\fnsymbol{CBtnc}}$\addtocounter{CBtnc}{1}\\*}{%
\centering{\small\mbox{}\protect\/{\protect\em School of Physics%
, The University of Melbourne, Parkville, Victoria 3052, Austral%
ia\protect\/}}\\}{\centering\protect\mbox{}\\(22 May 1995)\\}\par
\vspace\baselineskip\begin{centre}{\small\bf Abstract}\end{centr%
e}\vspace{-1.5ex}\vspace{-0.75\baselineskip}\par\setlength{\Csave
}{\parskip}\begin{quote}\setlength{\parskip}{0.5\baselineskip}%
\small\noindent In this short note, I point out that \mbox{$[p,q%
]\neq i\hbar$}, contrary to the original claims of \protect\ref{%
au:Born1925}, and \protect\ref{au:Dirac1926}. Rather, $[p,q]$ is
equal to something that is \mbox{}\protect\/{\protect\em infinit%
esimally different\protect\/} from $i\hbar$. While this differen%
ce is usually harmless, it does provide the solution of the Born%
--Jordan ``trace paradox'' of $[p,q]$. More recently, subtleties
of a very similar form have been found to be of fundamental impo%
rtance in quantum field theory. \end{quote}\setlength{\parskip}{%
\Csave}\protect\mbox{}\par\typeout{Bold latin symbols option fil%
e <1 March 1995>}\par\typeout{Bold greek symbols option file <1
March 1995>}\par When \protect\ref{au:Born1925}~[\ref{cit:Born19%
25}] and \protect\ref{au:Dirac1926}~[\ref{cit:Dirac1926}] discov%
ered the relationship \mbox{$[p,q]=i\hbar$}, it was a turning po%
int in physics. Classically, physical quantities had always been
assumed to \mbox{}\protect\/{\protect\em commute\protect\/}; qua%
ntum mechanics was born when this assumption was discarded. Matr%
ix mechanics reflects this non-commutativity by representing qua%
ntities such as $p$ and $q$ by \mbox{}\protect\/{\protect\em mat%
rices\protect\/}; wave mechanics does likewise by considering th%
em to be \mbox{}\protect\/{\protect\em operators\protect\/}; and
Dirac's c-number and q-number formulation simply takes non-commu%
tativity as the starting point.\par Born and Jordan obtained
\mbox{$[p,q]=i\hbar$} by means of arguments based on the corresp%
ondence principle; and Dirac obtained it by his Poisson bracket
ansatz. Let us review its standard wave-mechanical derivation. I%
n the $q$-representation, the state vector \mbox{$\protect\left|%
\psi\protect\right\rangle$} is a function $\psi(q)$, the operato%
r $q$ is simply multiplication by $q$, and the operator $p$ is d%
efined as $i\hbar\,\partial/\partial q$. Thus the identity \mbox
{$[p,q]=i\hbar$} is just a scaling by the factor $i\hbar$ of the
identity \setcounter{Ceqindent}{0}\protect\begin{eqnarray}%
\protect\left.\protect\begin{array}{rcl}\protect\displaystyle[%
\partial_q,q]=1,\setlength{\Cscr}{\value{CEht}\Ctenthex}%
\addtolength{\Cscr}{-1.0ex}\protect\raisebox{0ex}[\value{CEht}%
\Ctenthex][\Cscr]{}\protect\end{array}\protect\right.\protect
\label{eq:DxXComm}\protect\end{eqnarray}\setcounter{CEht}{10}whe%
re I am using the notation $\partial_q$ to denote $\partial/%
\partial q$. The meaning of (\protect\ref{eq:DxXComm}) is made m%
ore explicit if we write in the implied function $\psi(q)$ on bo%
th sides: \setcounter{Ceqindent}{0}\protect\begin{eqnarray}[%
\partial_q,q]\psi(q)=\psi(q).\protect\nonumber\setlength{\Cscr}{%
\value{CEht}\Ctenthex}\addtolength{\Cscr}{-1.0ex}\protect
\raisebox{0ex}[\value{CEht}\Ctenthex][\Cscr]{}\protect\end{eqnar%
ray}\setcounter{CEht}{10}It is straightforward to multiply out t%
he left-hand side, and use the product rule to obtain the right-%
hand side: \setcounter{Ceqindent}{0}\protect\begin{eqnarray}[%
\partial_q,q]\psi(q)\hspace{-1.3ex}&\displaystyle\equiv&\hspace{%
-1.3ex}\partial_qq\psi(q)-q\partial_q\psi(q)\protect\nonumber
\setlength{\Cscr}{\value{CEht}\Ctenthex}\addtolength{\Cscr}{-1.0%
ex}\protect\raisebox{0ex}[\value{CEht}\Ctenthex][\Cscr]{}\\[0ex]%
\protect\displaystyle\hspace{-1.3ex}&\displaystyle=&\hspace{-1.3%
ex}\psi(q)+q\partial_q\psi(q)-q\partial_q\psi(q)\protect\nonumber
\setlength{\Cscr}{\value{CEht}\Ctenthex}\addtolength{\Cscr}{-1.0%
ex}\protect\raisebox{0ex}[\value{CEht}\Ctenthex][\Cscr]{}\\[0ex]%
\protect\displaystyle\hspace{-1.3ex}&\displaystyle=&\hspace{-1.3%
ex}\psi(q).\protect\nonumber\setlength{\Cscr}{\value{CEht}%
\Ctenthex}\addtolength{\Cscr}{-1.0ex}\protect\raisebox{0ex}[%
\value{CEht}\Ctenthex][\Cscr]{}\protect\end{eqnarray}\setcounter
{CEht}{10}\par A problem, however, arises when we want to make t%
he transition to the \mbox{}\protect\/{\protect\em matrix mechan%
ical\protect\/} formulation of quantum mechanics. In this formul%
ation, the state vector \mbox{$\protect\left|\psi\protect\right
\rangle$} is written as a \mbox{}\protect\/{\protect\em column v%
ector\protect\/}. Physically observable quantities, such as $p$
and $q$, must be represented by \mbox{}\protect\/{\protect\em He%
rmitian matrices\protect\/}. The equivalent of the $q$-represent%
ation of wave mechanics is obtained by taking the vertical posit%
ion in the column vector as a linear function of the value $q$.
Since $q$ is a continuous variable, and the rows of a column vec%
tor are discrete, we must consider the \mbox{}\protect\/{\protect
\em limit\protect\/} of a sequence of discrete matrix representa%
tions, of ever increasing dimension, such that the positions in
the column vector for \mbox{$\protect\left|\psi\protect\right
\rangle$} ``fill in'' the domain of $q$ more and more densely, s%
o that in the limit of an infinite-dimensional matrix they form
a continuum. Let us label the rows in such a way that the ``midd%
le one'' has the index value $n=0$; the rows above are rows $n=-%
1$, $-2$, $-3$, \ldots, and those below are $n=+1$, $+2$, $+3$,
\ldots. Let us then deem that row~0 is to represent the origin o%
f the $q$ co\-ordinate, $q=0$. Then the relationship between $q$
and $n$ is of the form \setcounter{Ceqindent}{0}\protect\begin{e%
qnarray}q=\ell n,\protect\nonumber\setlength{\Cscr}{\value{CEht}%
\Ctenthex}\addtolength{\Cscr}{-1.0ex}\protect\raisebox{0ex}[%
\value{CEht}\Ctenthex][\Cscr]{}\protect\end{eqnarray}\setcounter
{CEht}{10}where $\ell$ is some length scale, that will shrink as
the dimension of the matrix representation is increased. (The pr%
ecise mathematical form of this ``shrinking rate'' does not need
to be known for our purposes). In other words, if we denote the
column vector representing \mbox{$\protect\left|\psi\protect
\right\rangle$} by the boldface symbol ${\protect\mbox{\protect
\boldmath{$\psi$}}}$, then we have \setcounter{Ceqindent}{0}%
\protect\begin{eqnarray}\protect\left.\protect\begin{array}{rcl}%
\protect\displaystyle\protect\left(\begin{array}{c}\vdots\\{%
\protect\mbox{\protect\boldmath{$\psi$}}}_{-2}\\{\protect\mbox{%
\protect\boldmath{$\psi$}}}_{-1}\\{\protect\mbox{\protect
\boldmath{$\psi$}}}_0\\{\protect\mbox{\protect\boldmath{$\psi$}}%
}_{+1}\\{\protect\mbox{\protect\boldmath{$\psi$}}}_{+2}\\\vdots
\end{array}\protect\right)\equiv\protect\left(\begin{array}{c}%
\vdots\\\psi(-2\ell)\\\psi(-\ell)\\\psi(0)\\\psi(+\ell)\\\psi(+2%
\ell)\\\vdots\end{array}\protect\right).\setlength{\Cscr}{\value
{CEht}\Ctenthex}\addtolength{\Cscr}{-1.0ex}\protect\raisebox{0ex%
}[\value{CEht}\Ctenthex][\Cscr]{}\protect\end{array}\protect
\right.\protect\label{eq:ColumnPsi}\protect\end{eqnarray}%
\setcounter{CEht}{10}\par Let us now construct the matrix ${%
\protect\mbox{\protect\boldmath{$q$}}}$ that represents the oper%
ator $q$. Clearly, the quantity $q\psi(q)$ is given in the matri%
x representation by \setcounter{Ceqindent}{0}\protect\begin{eqna%
rray}{\protect\mbox{\protect\boldmath{$q$}}}{\protect\mbox{%
\protect\boldmath{$\psi$}}}=\protect\left(\begin{array}{c}\vdots
\\-2\ell\,\psi(-2\ell)\\-\ell\,\psi(-\ell)\\0\\+\ell\,\psi(+\ell
)\\+2\ell\,\psi(+2\ell)\\\vdots\end{array}\protect\right);%
\protect\nonumber\setlength{\Cscr}{\value{CEht}\Ctenthex}%
\addtolength{\Cscr}{-1.0ex}\protect\raisebox{0ex}[\value{CEht}%
\Ctenthex][\Cscr]{}\protect\end{eqnarray}\setcounter{CEht}{10}fr%
om the identity (\protect\ref{eq:ColumnPsi}), it is then clear t%
hat \setcounter{Ceqindent}{0}\protect\begin{eqnarray}{\protect
\mbox{\protect\boldmath{$q$}}}=\ell\protect\left(\begin{array}{c%
cccccc}\!\raisebox{0ex}[1ex][0ex]{$\ddots$}\!&\cdot&\cdot&\cdot&%
\cdot&\cdot&\cdot\\\cdot&-2&\cdot&\cdot&\cdot&\cdot&\cdot\\\cdot
&\cdot&-1&\cdot&\cdot&\cdot&\cdot\\\cdot&\cdot&\cdot&\;0\;&\cdot
&\cdot&\cdot\\\cdot&\cdot&\cdot&\cdot&+1&\cdot&\cdot\\\cdot&\cdot
&\cdot&\cdot&\cdot&+2&\cdot\\\cdot&\cdot&\cdot&\cdot&\cdot&\cdot
&\!\raisebox{0ex}[1ex][0ex]{$\ddots$}\!\end{array}\protect\right
)\protect\nonumber\setlength{\Cscr}{\value{CEht}\Ctenthex}%
\addtolength{\Cscr}{-1.0ex}\protect\raisebox{0ex}[\value{CEht}%
\Ctenthex][\Cscr]{}\protect\end{eqnarray}\setcounter{CEht}{10}is
the matrix corresponding to the operator ${\protect\mbox{\protect
\boldmath{$q$}}}$, where dots indicate zero entries in the matri%
x.\par Constructing a matrix ${\protect\mbox{\protect\boldmath{$%
p$}}}$ to represent the operator $p$ is a little more subtle. Cl%
early, it relies on us devising a suitable matrix operator that
is equivalent to the operator $\partial_q$, in the limit of an i%
nfinite dimensional matrix. Now, since $p$ is postulated to be a%
n observable quantity, the matrix ${\protect\mbox{\protect
\boldmath{$p$}}}$ must be Hermitian; and since by definition $p%
\equiv i\hbar\partial_q$, then it follows that the matrix repres%
entation ${\protect\mbox{\protect\boldmath{$\partial$}}}_q$ of $%
\partial_q$ must be \mbox{}\protect\/{\protect\em anti-\protect
\/}Hermitian. Furthermore, since the derivative $\partial_q$ of
any \mbox{}\protect\/{\protect\em real\protect\/} function $\psi
(q)$ must itself be real, and since the definition of ${\protect
\mbox{\protect\boldmath{$\partial$}}}_q$ cannot depend on whethe%
r the function we apply it to is real or complex, then it follow%
s that ${\protect\mbox{\protect\boldmath{$\partial$}}}_q$ must i%
n full generality be a real matrix. Taken together, these two co%
nsiderations already tell us that ${\protect\mbox{\protect
\boldmath{$\partial$}}}_q$ must be a \mbox{}\protect\/{\protect
\em real, antisymmetric\protect\/} matrix. To find its exact for%
m, let us consider the meaning of the derivative $\partial_q$ fr%
om first principles: for a function $\psi(q)$, \setcounter{Ceqin%
dent}{0}\protect\begin{eqnarray}\protect\left.\protect\begin{arr%
ay}{rcl}\protect\displaystyle\partial_{q'}\psi(q')\setcounter{Cb%
scurr}{20}\setlength{\Cscr}{\value{Cbscurr}\Ctenthex}\addtolength
{\Cscr}{-1.0ex}\protect\raisebox{0ex}[\value{Cbscurr}\Ctenthex][%
\Cscr]{}\hspace{-0.05ex}{\protect\left|\setlength{\Cscr}{\value{%
Cbscurr}\Ctenthex}\addtolength{\Cscr}{-1.0ex}\protect\raisebox{0%
ex}[\value{Cbscurr}\Ctenthex][\Cscr]{}\protect\right.}\hspace{-0%
.25ex}\setlength{\Cscr}{\value{Cbscurr}\Ctenthex}\addtolength{%
\Cscr}{-1.0ex}\protect\raisebox{0ex}[\value{Cbscurr}\Ctenthex][%
\Cscr]{}_{q'=q}\equiv\lim_{\varepsilon,\varepsilon'\rightarrow0}%
\mbox{$\protect\displaystyle\protect\frac{\psi(q+\varepsilon)-%
\psi(q-\varepsilon')}{\varepsilon+\varepsilon'}$},\setlength{%
\Cscr}{\value{CEht}\Ctenthex}\addtolength{\Cscr}{-1.0ex}\protect
\raisebox{0ex}[\value{CEht}\Ctenthex][\Cscr]{}\protect\end{array%
}\protect\right.\protect\label{eq:DerivFirstPrinc}\protect\end{e%
qnarray}\setcounter{CEht}{10}where $\varepsilon$ and $\varepsilon
'$ are real numbers greater than zero. Now, in the matrix repres%
entation, for a finite dimension, we do not have positions that
are infinitesimally close to a given $q_n\equiv n\ell$; rather,
the closest we can get are the two points $q_{n+1}=(n+1)\ell$ an%
d $q_{n-1}=(n-1)\ell$. However, in the limit of an infinite-dime%
nsional matrix representation, these two points will shrink arou%
nd the point $q_n$ in the way we desire. Moreover, we already kn%
ow that we cannot use the point $q_n$ itself in the definition o%
f $\partial_q$, since the matrix ${\protect\mbox{\protect
\boldmath{$\partial$}}}_q$ must be antisymmetric, which means th%
at the diagonal elements must vanish. The best that we can do is
therefore \setcounter{Ceqindent}{0}\protect\begin{eqnarray}%
\protect\left.\protect\begin{array}{rcl}\protect\displaystyle({%
\protect\mbox{\protect\boldmath{$\partial$}}}_q{\protect\mbox{%
\protect\boldmath{$\psi$}}})_n\equiv\mbox{$\protect\displaystyle
\protect\frac{{\protect\mbox{\protect\boldmath{$\psi$}}}_{n+1}-{%
\protect\mbox{\protect\boldmath{$\psi$}}}_{n-1}}{2\ell}$},%
\setlength{\Cscr}{\value{CEht}\Ctenthex}\addtolength{\Cscr}{-1.0%
ex}\protect\raisebox{0ex}[\value{CEht}\Ctenthex][\Cscr]{}\protect
\end{array}\protect\right.\protect\label{eq:DerivMatrix}\protect
\end{eqnarray}\setcounter{CEht}{10}which is equivalent to (%
\protect\ref{eq:DerivFirstPrinc}), with $\varepsilon=\varepsilon
'$, in the limit of an infinite-dimensional matrix. This then im%
ples that \setcounter{Ceqindent}{0}\protect\begin{eqnarray}%
\protect\left.\protect\begin{array}{rcl}\protect\displaystyle{%
\protect\mbox{\protect\boldmath{$\partial$}}}_q=\mbox{$\protect
\displaystyle\protect\frac{1}{2\ell}$}\protect\left(\begin{array%
}{ccccccc}\!\raisebox{0ex}[1ex][0ex]{$\ddots$}\!&\!\raisebox{0ex%
}[1ex][0ex]{$\ddots$}\!&\cdot&\cdot&\cdot&\cdot&\cdot\\\!%
\raisebox{0ex}[1ex][0ex]{$\ddots$}\!&0&+1&\cdot&\cdot&\cdot&\cdot
\\\cdot&-1&0&+1&\cdot&\cdot&\cdot\\\cdot&\cdot&-1&0&+1&\cdot&%
\cdot\\\cdot&\cdot&\cdot&-1&0&+1&\cdot\\\cdot&\cdot&\cdot&\cdot&%
-1&0&\!\raisebox{0ex}[1ex][0ex]{$\ddots$}\!\\\cdot&\cdot&\cdot&%
\cdot&\cdot&\!\raisebox{0ex}[1ex][0ex]{$\ddots$}\!&\!\raisebox{0%
ex}[1ex][0ex]{$\ddots$}\!\end{array}\protect\right),\setlength{%
\Cscr}{\value{CEht}\Ctenthex}\addtolength{\Cscr}{-1.0ex}\protect
\raisebox{0ex}[\value{CEht}\Ctenthex][\Cscr]{}\protect\end{array%
}\protect\right.\protect\label{eq:MatrixPardQ}\protect\end{eqnar%
ray}\setcounter{CEht}{10}as can be verified directly by multiply%
ing (\protect\ref{eq:MatrixPardQ}) by (\protect\ref{eq:ColumnPsi%
}). In other words, the matrix ${\protect\mbox{\protect\boldmath
{$p$}}}$ is given by \setcounter{Ceqindent}{0}\protect\begin{eqn%
array}\protect\left.\protect\begin{array}{rcl}\protect
\displaystyle{\protect\mbox{\protect\boldmath{$p$}}}=\mbox{$%
\protect\displaystyle\protect\frac{\hbar}{2\ell}$}\protect\left(%
\begin{array}{ccccccc}\!\raisebox{0ex}[1ex][0ex]{$\ddots$}\!&\!%
\raisebox{0ex}[1ex][0ex]{$\ddots$}\!&\cdot&\cdot&\cdot&\cdot&%
\cdot\\\!\raisebox{0ex}[1ex][0ex]{$\ddots$}\!&0&+i&\cdot&\cdot&%
\cdot&\cdot\\\cdot&-i&0&+i&\cdot&\cdot&\cdot\\\cdot&\cdot&-i&0&+%
i&\cdot&\cdot\\\cdot&\cdot&\cdot&-i&0&+i&\cdot\\\cdot&\cdot&\cdot
&\cdot&-i&0&\!\raisebox{0ex}[1ex][0ex]{$\ddots$}\!\\\cdot&\cdot&%
\cdot&\cdot&\cdot&\!\raisebox{0ex}[1ex][0ex]{$\ddots$}\!&\!%
\raisebox{0ex}[1ex][0ex]{$\ddots$}\!\end{array}\protect\right),%
\setlength{\Cscr}{\value{CEht}\Ctenthex}\addtolength{\Cscr}{-1.0%
ex}\protect\raisebox{0ex}[\value{CEht}\Ctenthex][\Cscr]{}\protect
\end{array}\protect\right.\protect\label{eq:MatrixP}\protect\end
{eqnarray}\setcounter{CEht}{10}which is clearly Hermitian, as re%
quired.\par We can now turn immediately to the issue raised by t%
he title of this note, by computing $[p,q]$ in the matrix repres%
entation above---namely, by computing the matrix commutator $[{%
\protect\mbox{\protect\boldmath{$p$}}},{\protect\mbox{\protect
\boldmath{$q$}}}]$. By multiplying out the matrices, it is easil%
y seen that \setcounter{Ceqindent}{0}\protect\begin{eqnarray}{%
\protect\mbox{\protect\boldmath{$p$}}}{\protect\mbox{\protect
\boldmath{$q$}}}=\mbox{$\protect\displaystyle\protect\frac{i\hbar
}{2}$}\protect\left(\begin{array}{ccccccc}\!\raisebox{0ex}[1ex][%
0ex]{$\ddots$}\!&\!\raisebox{0ex}[1ex][0ex]{$\ddots$}\!&\cdot&%
\cdot&\cdot&\cdot&\cdot\\\!\raisebox{0ex}[1ex][0ex]{$\ddots$}\!&%
0&-1&\cdot&\cdot&\cdot&\cdot\\\cdot&+2&0&0&\cdot&\cdot&\cdot\\%
\cdot&\cdot&+1&\;0\;&+1&\cdot&\cdot\\\cdot&\cdot&\cdot&0&0&+2&%
\cdot\\\cdot&\cdot&\cdot&\cdot&-1&0&\!\raisebox{0ex}[1ex][0ex]{$%
\ddots$}\!\\\cdot&\cdot&\cdot&\cdot&\cdot&\!\raisebox{0ex}[1ex][%
0ex]{$\ddots$}\!&\!\raisebox{0ex}[1ex][0ex]{$\ddots$}\!\end{arra%
y}\protect\right)\protect\nonumber\setlength{\Cscr}{\value{CEht}%
\Ctenthex}\addtolength{\Cscr}{-1.0ex}\protect\raisebox{0ex}[%
\value{CEht}\Ctenthex][\Cscr]{}\protect\end{eqnarray}\setcounter
{CEht}{10}and \setcounter{Ceqindent}{0}\protect\begin{eqnarray}{%
\protect\mbox{\protect\boldmath{$q$}}}{\protect\mbox{\protect
\boldmath{$p$}}}=\mbox{$\protect\displaystyle\protect\frac{i\hbar
}{2}$}\protect\left(\begin{array}{ccccccc}\!\raisebox{0ex}[1ex][%
0ex]{$\ddots$}\!&\!\raisebox{0ex}[1ex][0ex]{$\ddots$}\!&\cdot&%
\cdot&\cdot&\cdot&\cdot\\\!\raisebox{0ex}[1ex][0ex]{$\ddots$}\!&%
0&-2&\cdot&\cdot&\cdot&\cdot\\\cdot&+1&0&-1&\cdot&\cdot&\cdot\\%
\cdot&\cdot&0&0&0&\cdot&\cdot\\\cdot&\cdot&\cdot&-1&0&+1&\cdot\\%
\cdot&\cdot&\cdot&\cdot&-2&0&\!\raisebox{0ex}[1ex][0ex]{$\ddots$%
}\!\\\cdot&\cdot&\cdot&\cdot&\cdot&\!\raisebox{0ex}[1ex][0ex]{$%
\ddots$}\!&\!\raisebox{0ex}[1ex][0ex]{$\ddots$}\!\end{array}%
\protect\right);\protect\nonumber\setlength{\Cscr}{\value{CEht}%
\Ctenthex}\addtolength{\Cscr}{-1.0ex}\protect\raisebox{0ex}[%
\value{CEht}\Ctenthex][\Cscr]{}\protect\end{eqnarray}\setcounter
{CEht}{10}we therefore find that \setcounter{Ceqindent}{0}%
\protect\begin{eqnarray}\protect\left.\protect\begin{array}{rcl}%
\protect\displaystyle[{\protect\mbox{\protect\boldmath{$p$}}},{%
\protect\mbox{\protect\boldmath{$q$}}}]=\mbox{$\protect
\displaystyle\protect\frac{i\hbar}{2}$}\protect\left(\begin{arra%
y}{ccccccc}\!\raisebox{0ex}[1ex][0ex]{$\ddots$}\!&\!\raisebox{0e%
x}[1ex][0ex]{$\ddots$}\!&\cdot&\cdot&\cdot&\cdot&\cdot\\\!%
\raisebox{0ex}[1ex][0ex]{$\ddots$}\!&\:0\:&\:1\:&\cdot&\cdot&%
\cdot&\cdot\\\cdot&\:1\:&\:0\:&\:1\:&\cdot&\cdot&\cdot\\\cdot&%
\cdot&\:1\:&\:0\:&\:1\:&\cdot&\cdot\\\cdot&\cdot&\cdot&\:1\:&\:0%
\:&\:1\:&\cdot\\\cdot&\cdot&\cdot&\cdot&\:1\:&\:0\:&\!\raisebox{%
0ex}[1ex][0ex]{$\ddots$}\!\\\cdot&\cdot&\cdot&\cdot&\cdot&\!%
\raisebox{0ex}[1ex][0ex]{$\ddots$}\!&\!\raisebox{0ex}[1ex][0ex]{%
$\ddots$}\!\end{array}\protect\right).\setlength{\Cscr}{\value{C%
Eht}\Ctenthex}\addtolength{\Cscr}{-1.0ex}\protect\raisebox{0ex}[%
\value{CEht}\Ctenthex][\Cscr]{}\protect\end{array}\protect\right
.\protect\label{eq:PQComm}\protect\end{eqnarray}\setcounter{CEht%
}{10}\mbox{}\protect\/{\protect\em Here is the subtlety.\protect
\/} The problem is that the matrix (\protect\ref{eq:PQComm}) is
\mbox{}\protect\/{\protect\em not\protect\/} equal to $i\hbar$ t%
imes the unit matrix $\mbox{\bf1}$, \setcounter{Ceqindent}{0}%
\protect\begin{eqnarray}\protect\left.\protect\begin{array}{rcl}%
\protect\displaystyle i\hbar\,\mbox{\bf1}\equiv\mbox{$\protect
\displaystyle\protect\frac{i\hbar}{2}$}\protect\left(\begin{arra%
y}{ccccccc}\!\raisebox{0ex}[1ex][0ex]{$\ddots$}\!&\cdot&\cdot&%
\cdot&\cdot&\cdot&\cdot\\\cdot&\:2\:&\cdot&\cdot&\cdot&\cdot&%
\cdot\\\cdot&\cdot&\:2\:&\cdot&\cdot&\cdot&\cdot\\\cdot&\cdot&%
\cdot&\:2\:&\cdot&\cdot&\cdot\\\cdot&\cdot&\cdot&\cdot&\:2\:&%
\cdot&\cdot\\\cdot&\cdot&\cdot&\cdot&\cdot&\:2\:&\cdot\\\cdot&%
\cdot&\cdot&\cdot&\cdot&\cdot&\!\raisebox{0ex}[1ex][0ex]{$\ddots
$}\!\end{array}\protect\right).\setlength{\Cscr}{\value{CEht}%
\Ctenthex}\addtolength{\Cscr}{-1.0ex}\protect\raisebox{0ex}[%
\value{CEht}\Ctenthex][\Cscr]{}\protect\end{array}\protect\right
.\protect\label{eq:Unit}\protect\end{eqnarray}\setcounter{CEht}{%
10}Rather, the matrix (\protect\ref{eq:PQComm}) is effectively o%
btained by taking each diagonal element of $2$ and ``splitting i%
t'' between the off-diagonals above and below. Thus we have prov%
ed the relation \setcounter{Ceqindent}{0}\protect\begin{eqnarray%
}\protect\left.\protect\begin{array}{rcl}\protect\displaystyle[{%
\protect\mbox{\protect\boldmath{$p$}}},{\protect\mbox{\protect
\boldmath{$q$}}}]\neq i\hbar\mbox{\bf1}\setlength{\Cscr}{\value{%
CEht}\Ctenthex}\addtolength{\Cscr}{-1.0ex}\protect\raisebox{0ex}%
[\value{CEht}\Ctenthex][\Cscr]{}\protect\end{array}\protect\right
.\protect\label{eq:MatrixNotEq}\protect\end{eqnarray}\setcounter
{CEht}{10}in the matrix representation of quantum mechanics, and
hence in full generality \setcounter{Ceqindent}{0}\protect\begin
{eqnarray}\protect\left.\protect\begin{array}{rcl}\protect
\displaystyle[p,q]\neq i\hbar,\setlength{\Cscr}{\value{CEht}%
\Ctenthex}\addtolength{\Cscr}{-1.0ex}\protect\raisebox{0ex}[%
\value{CEht}\Ctenthex][\Cscr]{}\protect\end{array}\protect\right
.\protect\label{eq:NotEq}\protect\end{eqnarray}\setcounter{CEht}%
{10}as I have claimed in the title of this note.\par The result
(\protect\ref{eq:NotEq}) might be somewhat disturbing. However,
in almost all cases, it is of academic interest only. The reason
is that using the matrix (\protect\ref{eq:Unit}) rather than (%
\protect\ref{eq:PQComm}), in any practical calculation, correspo%
nds to the replacement \setcounter{Ceqindent}{0}\protect\begin{e%
qnarray}\lim_{\ell\rightarrow0}\mbox{$\protect\displaystyle
\protect\frac{\psi(q-\ell)+\psi(q+\ell)}{2}$}\rightarrow\psi(q),%
\protect\nonumber\setlength{\Cscr}{\value{CEht}\Ctenthex}%
\addtolength{\Cscr}{-1.0ex}\protect\raisebox{0ex}[\value{CEht}%
\Ctenthex][\Cscr]{}\protect\end{eqnarray}\setcounter{CEht}{10}wh%
ich is arguably harmless for any reasonable $\psi(q)$. In fact,
we can obtain exactly the same result \mbox{$[p,q]\neq i\hbar$}
using the \mbox{}\protect\/{\protect\em wave mechanical\protect
\/} representation, if we treat the operation of differentiation
more carefully, rather than by simply using the product rule. If
we write down the wave-mechanical equivalent of the Hermitian (%
\mbox{}\protect\/{\protect\em i.e.\protect\/}, symmetrical) defi%
nition (\protect\ref{eq:DerivMatrix}), namely, the symmetrical v%
ersion of (\protect\ref{eq:DerivFirstPrinc}), \setcounter{Ceqind%
ent}{0}\protect\begin{eqnarray}\partial_{q'}\psi(q')\setcounter{%
Cbscurr}{20}\setlength{\Cscr}{\value{Cbscurr}\Ctenthex}%
\addtolength{\Cscr}{-1.0ex}\protect\raisebox{0ex}[\value{Cbscurr%
}\Ctenthex][\Cscr]{}\hspace{-0.05ex}{\protect\left|\setlength{%
\Cscr}{\value{Cbscurr}\Ctenthex}\addtolength{\Cscr}{-1.0ex}%
\protect\raisebox{0ex}[\value{Cbscurr}\Ctenthex][\Cscr]{}\protect
\right.}\hspace{-0.25ex}\setlength{\Cscr}{\value{Cbscurr}%
\Ctenthex}\addtolength{\Cscr}{-1.0ex}\protect\raisebox{0ex}[%
\value{Cbscurr}\Ctenthex][\Cscr]{}_{q'=q}\equiv\lim_{\varepsilon
\rightarrow0}\mbox{$\protect\displaystyle\protect\frac{\psi(q+%
\varepsilon)-\psi(q-\varepsilon)}{2\varepsilon}$},\protect
\nonumber\setlength{\Cscr}{\value{CEht}\Ctenthex}\addtolength{%
\Cscr}{-1.0ex}\protect\raisebox{0ex}[\value{CEht}\Ctenthex][\Cscr
]{}\protect\end{eqnarray}\setcounter{CEht}{10}then we find that
\setcounter{Ceqindent}{0}\protect\begin{eqnarray}\setcounter{CEh%
t}{30}[\partial_q,q]\psi(q)\hspace{-1.3ex}&\displaystyle\equiv&%
\hspace{-1.3ex}\partial_qq\psi(q)-q\partial_q\psi(q)\protect
\nonumber\setlength{\Cscr}{\value{CEht}\Ctenthex}\addtolength{%
\Cscr}{-1.0ex}\protect\raisebox{0ex}[\value{CEht}\Ctenthex][\Cscr
]{}\\[0ex]\protect\displaystyle\hspace{-1.3ex}&\displaystyle
\equiv&\hspace{-1.3ex}\lim_{\varepsilon\rightarrow0}\protect\left
\{\mbox{$\protect\displaystyle\protect\frac{(q+\varepsilon)\,\psi
(q+\varepsilon)-(q-\varepsilon)\,\psi(q-\varepsilon)}{2%
\varepsilon}$}-q\,\mbox{$\protect\displaystyle\protect\frac{\psi
(q+\varepsilon)-\psi(q-\varepsilon)}{2\varepsilon}$}\protect
\right\}\protect\nonumber\setlength{\Cscr}{\value{CEht}\Ctenthex
}\addtolength{\Cscr}{-1.0ex}\protect\raisebox{0ex}[\value{CEht}%
\Ctenthex][\Cscr]{}\\[0ex]\protect\displaystyle\hspace{-1.3ex}&%
\displaystyle\equiv&\hspace{-1.3ex}\lim_{\varepsilon\rightarrow0%
}\mbox{$\protect\displaystyle\protect\frac{\psi(q+\varepsilon)+%
\psi(q-\varepsilon)}{2}$},\protect\nonumber\setlength{\Cscr}{%
\value{CEht}\Ctenthex}\addtolength{\Cscr}{-1.0ex}\protect
\raisebox{0ex}[\value{CEht}\Ctenthex][\Cscr]{}\protect\end{eqnar%
ray}\setcounter{CEht}{10}in agreement with the matrix mechanical
result.\par It might seem that claiming that \mbox{$[p,q]\neq i%
\hbar$} is a pedantry. After all, when would shifting the argume%
nt $q$ by an infinitesimal amount, or shifting by one row or col%
umn in an infinite-dimensional matrix representation, make any d%
ifference? There is at least one situation that I am aware of in
which this change \mbox{}\protect\/{\protect\em does\protect\/}
make a difference: whenever the \mbox{}\protect\/{\protect\em tr%
ace\protect\/} of the matrix is taken. For example, the Born and
Jordan's~[\ref{cit:Born1925}] well-known ``trace paradox'' of $[%
p,q]$ points out the following: since \setcounter{Ceqindent}{0}%
\protect\begin{eqnarray}\mbox{Tr}({\protect\mbox{\protect
\boldmath{$A$}}}{\protect\mbox{\protect\boldmath{$B$}}})\equiv
\mbox{Tr}({\protect\mbox{\protect\boldmath{$B$}}}{\protect\mbox{%
\protect\boldmath{$A$}}})\protect\nonumber\setlength{\Cscr}{%
\value{CEht}\Ctenthex}\addtolength{\Cscr}{-1.0ex}\protect
\raisebox{0ex}[\value{CEht}\Ctenthex][\Cscr]{}\protect\end{eqnar%
ray}\setcounter{CEht}{10}for any finite matrices ${\protect\mbox
{\protect\boldmath{$A$}}}$ and ${\protect\mbox{\protect\boldmath
{$B$}}}$, then in the finite-dimensional case we must have
\setcounter{Ceqindent}{0}\protect\begin{eqnarray}\mbox{Tr}[{%
\protect\mbox{\protect\boldmath{$p$}}},{\protect\mbox{\protect
\boldmath{$q$}}}]\equiv\mbox{Tr}({\protect\mbox{\protect\boldmath
{$p$}}}{\protect\mbox{\protect\boldmath{$q$}}}-{\protect\mbox{%
\protect\boldmath{$q$}}}{\protect\mbox{\protect\boldmath{$p$}}})%
\equiv0.\protect\nonumber\setlength{\Cscr}{\value{CEht}\Ctenthex
}\addtolength{\Cscr}{-1.0ex}\protect\raisebox{0ex}[\value{CEht}%
\Ctenthex][\Cscr]{}\protect\end{eqnarray}\setcounter{CEht}{10}Bu%
t if \mbox{$[p,q]=i\hbar$} were to hold true, then we would need
to have \setcounter{Ceqindent}{0}\protect\begin{eqnarray}\mbox{T%
r}[{\protect\mbox{\protect\boldmath{$p$}}},{\protect\mbox{%
\protect\boldmath{$q$}}}]=i\hbar\,\mbox{Tr}(\mbox{\bf1})=i\hbar
\,D,\protect\nonumber\setlength{\Cscr}{\value{CEht}\Ctenthex}%
\addtolength{\Cscr}{-1.0ex}\protect\raisebox{0ex}[\value{CEht}%
\Ctenthex][\Cscr]{}\protect\end{eqnarray}\setcounter{CEht}{10}wh%
ere $D$ is the dimension of the matrix representation, which, ra%
ther than vanishing, approaches infinity in the infinite-dimensi%
onal limit! I emphasise that \mbox{}\protect\/{\protect\em this
is a fallacy\protect\/}; it is the matrix (\protect\ref{eq:PQCom%
m}) that must be used, \mbox{}\protect\/{\protect\em not\protect
\/} the identity matrix (\protect\ref{eq:Unit}). And of course t%
he matrix (\protect\ref{eq:PQComm}) is identically \mbox{}%
\protect\/{\protect\em traceless\protect\/}; hence, the Born--Jo%
rdan ``trace paradox'' of $[p,q]$ is due to the incorrect assump%
tion that \mbox{$[p,q]=i\hbar$}, whereas at the level of individ%
ual rows and columns of the matrix representation it fails.\par I%
t might be claimed that this simply shows that one cannot take i%
nfinite-dimensional matrix mechanics to be the infinite-dimensio%
n limit of finite-dimensional matrix mechanics. But then what wo%
uld this ``matrix mechanics'' have to do with matrices as we kno%
w them? Moreover, it is generally believed that the correct way
of dealing with infinities, or infinitesimals, in physical probl%
ems is to take them to be the limit of \mbox{}\protect\/{\protect
\em finite\protect\/} quantities. Surely, then, it is better to
modify the \mbox{}\protect\/{\protect\em postulate\protect\/} of
\mbox{$[p,q]=i\hbar$}\, by an infinitesimal amount, rather than
remove all chance of using a well-defined limiting procedure?\par
 Furthermore, this ability of the trace---to be able to yield an
answer that is either zero or infinite, depending on how careles%
sly one defines one's matrix quantities---turns out to be more i%
mportant to real-world calculations than one might na\-\"\i vely
think. In quantum field theory, the effect of effectively ``incl%
uding the diagonal terms'' in the time-ordered product operation%
, when in fact they should \mbox{}\protect\/{\protect\em not%
\protect\/} be included, leads to a drastic and fundamental chan%
ge in the predictions of calculations involving loop diagrams. T%
his has has been pointed out several times in the past two decad%
es, but has not gained much attention; we shall be providing a f%
ull and explicit description of these developments shortly~[\ref
{cit:Costella1995}].\par\vspace{1.5\baselineskip}\par{\centering
\bf Acknowledgments\\*[0.5\baselineskip]}\protect\indent Helpful
discussions with T.~D~Kieu, B.~H.~J.~McKellar, A.~A.~Rawlinson,
M.~J.~Thomson and G.~J.~Stephenson,~Jr.\ are gratefully acknowle%
dged. This work was supported in part by the Australian Research
Council.\par\vspace{1.5\baselineskip}\par{\centering\bf Referenc%
es\\*[0.5\baselineskip]}\protect\mbox{}\vspace{-\baselineskip}%
\vspace{-2ex}\settowidth\CGDnum{[\ref{citlast}]}\setlength{%
\CGDtext}{\textwidth}\addtolength{\CGDtext}{-\CGDnum}\begin{list%
}{Error!}{\setlength{\labelwidth}{\CGDnum}\setlength{\labelsep}{%
0.75ex}\setlength{\leftmargin}{\labelwidth}\addtolength{%
\leftmargin}{\labelsep}\setlength{\rightmargin}{0ex}\setlength{%
\itemsep}{0ex}\setlength{\parsep}{0ex}}\protect\frenchspacing
\setcounter{CBtnc}{1}\item[{\hfill\makebox[0ex][r]{\raisebox{0ex%
}[1ex][0ex]{$^{\mbox{$\fnsymbol{CBtnc}$}}$}}}]\addtocounter{CBtn%
c}{1}jpc@physics.unimelb.edu.au; http:/$\!$/www.ph.unimelb.edu.a%
u/$\sim$jpc/homepage.htm.\nopagebreak\addtocounter{CBcit}{1}\item
[\hfill{[}\arabic{CBcit}{]}]\renewcommand\theCscr{\arabic{CBcit}%
}\protect\refstepcounter{Cscr}\protect\label{cit:Born1925}M.~Bor%
n and P.~Jordan, \renewcommand\theCscr{Born and Jordan}\protect
\refstepcounter{Cscr}\protect\label{au:Born1925}\renewcommand
\theCscr{1925}\protect\refstepcounter{Cscr}\protect\label{yr:Bor%
n1925}\mbox{}\protect\/{\protect\em Z. Phys.\protect\/} {\bf34}
(1925) 858.\addtocounter{CBcit}{1}\item[\hfill{[}\arabic{CBcit}{%
]}]\renewcommand\theCscr{\arabic{CBcit}}\protect\refstepcounter{%
Cscr}\protect\label{cit:Dirac1926}P.~A.~M.~Dirac, \renewcommand
\theCscr{Dirac}\protect\refstepcounter{Cscr}\protect\label{au:Di%
rac1926}\renewcommand\theCscr{1926}\protect\refstepcounter{Cscr}%
\protect\label{yr:Dirac1926}\mbox{}\protect\/{\protect\em Proc.
Roy. Soc. London~A\protect\/} {\bf109} (1926) 642.\addtocounter{%
CBcit}{1}\item[\hfill{[}\arabic{CBcit}{]}]\renewcommand\theCscr{%
\arabic{CBcit}}\protect\refstepcounter{Cscr}\protect\label{cit:C%
ostella1995}J.~P.~Costella, T.~D.~Kieu, B.~H.~J.~McKellar, A.~A.%
 ~Rawlinson, M.~J.~Thomson and G.~J.~Stephenson, Jr.,
\renewcommand\theCscr{Costella, Kieu, McKellar, Rawlinson, Thoms%
on and Stephenson, Jr.}\protect\refstepcounter{Cscr}\protect
\label{au:Costella1995}\renewcommand\theCscr{1995}\protect
\refstepcounter{Cscr}\protect\label{yr:Costella1995}in preparati%
on.\renewcommand\theCscr{\arabic{CBcit}}\protect\refstepcounter{%
Cscr}\protect\label{citlast}\settowidth\Cscr{~[\ref{cit:Costella%
1995}]}\end{list}\par\end{document}